# The Formation of Galaxies: A Challenge for Supercomputers – a Simple Task for GRAPE ?

Matthias Steinmetz

Max–Planck–Institut für Astrophysik, Postfach 1523, 85740 Garching (FRG)
e-mail: mhs@MPA-Garching.MPG.DE

**Abstract.** We present numerical simulations of galaxy formation, one of the most challenging problems in computational astrophysics. The key point in such simulations is the efficient solution of the N–body problem. If the gas of a galaxy is treated by means of smoothed particle hydrodynamics (SPH), the hydrodynamic equations can be reduced to a form similar to that of the N–body problem. A straightforward implementation requires a computational effort $\propto N^2$, making it prohibitively expensive to simulate systems larger than $10^5$ particles even on the largest available supercomputers.

After a description of the physical and numerical problems, we shortly review the standard numerical methods to tackle these problems and discuss their advantages and drawbacks. We also present a completely different approach to perform such simulations using a workstation in combination with the special purpose hardware GRAPE. After a discussion of the main features of GRAPE, we present a simple implementation of a SPH–N-body code on such a configuration. Comparing results and performance of these two approaches, we show, that with an investment of US $ 50000, the problem can be solved up to 5 times faster than on a CRAY YMP.

## 1 Introduction

During the last decade one of the most active fields of astrophysics has been the study of the formation of galaxies. It is commonly believed, that up to 95% of the matter of the universe is composed of dark matter, which is probably of non–baryonic origin and interacts mainly via gravitation. According to a widely accepted idea galaxies form by the collapse of gravitationally unstable primordial density fluctuations. The evolution of the gravitationally dominant dark matter is treated by N–body techniques. During the last years, one has begun to add gasdynamics to the simulations to mimic the evolution of the baryons. In some simulations, stars and galaxies are formed out of the collapsing gas, which are again treated by N–body techniques. The grand computational challenge of such simulations is twofold: Firstly, one can show [1], that the gravitational collapse must proceed anisotropically, i.e., three–dimensional simulations are necessary. Secondly, very different length scales are involved, starting from 1 pc ($\approx 3.1\ 10^{13}$ km), which is the size of a typical star forming region, up to several hundred Mpc, a volume which can be considered to be a representative piece of our universe. This range of scales must be compared with the largest simulations feasible nowadays, which are performed on a grid of about $300^3$ zones. Besides classical finite–difference methods, a completely different approach to solve the hydrodynamic equations is used in the astrophysical community: *Smoothed Particle Hydrodynamics* (SPH, [2]). Its main advantage is to be a free Lagrangean method. This makes it optimally suited for highly irregular clustered systems like galaxies. Although it is still to prove mathematically that the SPH equations converge to the hydrodynamical fluid equations, a series of test calculations has shown that the quality of SPH results can compete with that of modern finite difference schemes [3], even with suprisingly small particle numbers!

## 2 Current techniques

The key problem to perform large scale computer simulations of structure formation is an efficient solution of the N–body problem, i.e., the calculation of the acceleration $\frac{d}{dt}\mathbf{v}_i$ of the particle $i$ due to the gravitational interaction with all other particles $j$ of the system:

$$\frac{d\mathbf{v}_i}{dt} = -G \sum_{j \neq i} \frac{m_j}{\left(r_{ij}^2 + s^2\right)^{3/2}} (\mathbf{r}_i - \mathbf{r}_j), \qquad (1)$$

with $r_{ij} = |\mathbf{r}_i - \mathbf{r}_j|$. $s$ is the so called softening parameter, which prevents the $1/r^2$ divergence of the force for $r \to 0$ and limits the resolution. Methods which directly solve the system (1) are called *Particle–Particle* (PP) methods [4]. Because gravity is a long range force, the computational effort to determine the force on all $N$ particles grows $\propto N^2$. This makes it prohibitively expensive to perform simulations involving much more than $10^4$ particles, even on the fastest available supercomputers. In SPH, the force law is of similar form, it is given by

$$\begin{aligned}
\langle \varrho(\mathbf{r}_i) \rangle &= \int d^3 r' \varrho(\mathbf{r}') W(\mathbf{r}_i - \mathbf{r}', h) \approx \sum m W(\mathbf{r}_i - \mathbf{r}_i, h) \\
\frac{d\mathbf{v}_i}{dt} &= \left.\frac{d\mathbf{v}_i}{dt}\right|_{\text{grav}} - \sum_j m_j \left(\frac{P_i}{\varrho_i} + \frac{P_j}{\varrho_j} + Q_{ij}\right) \nabla_i W(r_{ij}, h) \qquad (2) \\
\frac{d\varepsilon_i}{dt} &= \sum_j m_j \left(\frac{P_i}{\varrho_i^2} + \frac{1}{2} Q_{ij}\right) (\mathbf{v}_i - \mathbf{v}_j) \cdot \nabla_i W(r_{ij}, h).
\end{aligned}$$

In this equations, $W$ is the interpolation kernel with a shape similar to a Gaussian. $h$ is the smoothing length. $d\mathbf{v}_i/dt|_{\text{grav}}$ is the gravitational acceleration according to Eq. (1). Besides the particle number, $h$ determines the resolution of the system. The pressure $P$, the internal energy $\varepsilon$ and the density $\varrho$ are related by an equation of state. $Q_{ij}$ is an artificial viscosity introduced to treat shock waves. Note, that as long as the kernel $W$ has compact support, all the corrections of Eqs (2) to Eq. (1) are of short range nature, i.e., the additional computational effort is only $\propto N$.

In the past, several techniques have been developed to circumvent the $N^2$ behaviour: Particle–mesh methods (PM) do not explicitly solve Eq. (1). Instead, the distribution of particles is assigned to a grid, the mass per zone defining a density. Via Fast Fourier Transform (FFT), Poisson's equation is solved on the grid. The forces are than interpolated to the particle position. The computational effort grows only $\propto N \log M$, where $M$ is the number of zones per dimension. Though PM schemes are very fast, they are only suitable for relatively homogeneous systems. The resolution of a PM calculation is determined by the size of a zone, and even the largest current PM simulations with $500^3$ zones and $250^3$ particles have only a very limited spatial resolution. One tries to circumvent these problems by the $P^3M$ (= PP PM) technique. Here, the long range forces are calculated via the PM method, but the short range forces exerted by particles in the same or in the neighbouring zones are treated by the PP technique. The main drawback of $P^3M$ is that for highly clumped structures, a large number of particles is placed within a few zones, and the PP part becomes computationally dominant. For very large simulations ($200^3$) particle numbers exceeding $10^4$ per zone are not atypical. The computational effort of $P^3M$ is difficult to estimate: for a homogeneous system it is $\propto N \log M$ as for PM, in the worst case, most of the paricles are located in a few cells, the performance is degraded to the $N^2$ behaviour of the PP method. Another approach is the tree algorithm [5], [6]. Here, the main idea is to group distant particles together and to approximate the force exerted by this group by that of one particle of the same mass. A tree data structure is used to systematically group particles together. Comparing the extension of the group with its distance to a specific particle, one can determine whether this force approximation is accurate enough, or whether the group has to be split into subgroups, for which the same procedure is applied. The result is, that instead of $N$ only $\propto \log N$ interactions are to be calculated for every particle. Therefore, the computational effort scales like $N \log N$. The performance of tree methods also decreases with increasing clumpiness, although much weaker than $P^3M$ methods. However, the construction of the tree causes some overhead, which may become critical in a multiple timestep scheme, if only the force for a few particles has to be calculated. Finally, tree algorithms are relatively complex and require a lot of memory.

There exist various implementations of SPH and N–body codes on vector computers, but only very few on massively parallel machines (e.g. [7], [8], [9], [10]). Therefore, only very vague statements can be made about their performance. To implement very efficient PP codes on vector or shared memory machines is quite easy, but on a distributed memory machine this task is not unproblematic. For PM and $P^3M$ the FFT part can be handled efficiently. However, the mesh assignment of particles and the force interpolation are difficult to vectorize and involve a lot of indirect addressing. In case of parallel machines it is difficult to achieve a good load balancing for the mesh asignment and interpolation step. The recursive structure of a tree algorithm is difficult to vectorize, a lot of indirect addressing combined with short vector lengths is involved. By area decomposition, it is possible to run a tree code on a distributed memory machine with high performance [7], but major changes to the standard tree code have to be made. The currently largest N-body simulations ($260^3$) were done with such a code [11]. However, we think it is difficult to get a good load balancing if a multiple time step scheme is used, which is essential for a good performance of a SPH code.

In summary, there exist different techniques to tackle the N–body system with good performance on current supercomputers, although it is difficult to come close to the peak performance. For a good compromise between speed and resolution, $P^3M$ and tree codes are the favourite choices. Simulations involving $300^3$ particles are the current limit. In combination with SPH, the respective numbers are much smaller and even the largest simulations involve only a few times $10^5$ particles. The main reasons are: (i) The force calculation becomes more expensive. (ii) The particles have an extension, which increases the number of short range force evaluations. (iii) The number of time steps is 10–100 times higher. The resources necessary to perform a typical simulation of the formation of galaxies (timings for a tree code on one CRAY YMP processor) are the following: A 4000 particle N–body simulation ($\approx 1000$ timesteps) requires 40 min. The same simulation using a two component system of 4000 SPH and 4000 N–body particles (15000 timesteps) requires about 10 h, and a simulation with 4000 SPH, 4000 N–body, and at the end about 25000 star particles (N–body) needs 60 hours. Simulations with $32^3$ gas and dark matter particles as performed by [12] have consumed more than 200 CRAY hours. Calculations with a million particles of two or three different species would consume several thousands of CRAY hours. Finally note, that although these calculations may be regarded as grand challenges for future generations of supercomputers, from the physical point of view they still have only a very limited resolution!

## 3  The GRAPE Project

Up to now, all methods are based on software development. A completely different approach was chosen by Sugimoto and collaborators at the University of Tokyo (for an overview see [13]): The calculation of one force interaction is a combination of a very few specific arithmetic operations: three differences, three squares, one sum, etc. Then, the inter particle forces have to be summed up. Furthermore, many of these operations do not depend on each other and can be done in a pipeline. Since the number of operations to get the total force on one particle is the same for every particle, one can parallelize it with a very good load balancing. Sugimoto et al. have designed a series of special purpose hardware boards GRAPE (GRAvity PipE) to calculate (1). Furthermore, a list of particles within a sphere of a given radius $h_i$ of particle $i$ is returned, too. This is very helpful for an implementation of SPH. The board is connected to a workstation via a VME interface. Libraries allow one to use GRAPE by FORTRAN or C subroutine calls. The prototype GRAPE1 reached 240 Mflops in 1989. Meanwhile there exist two series of boards: the odd numbers (GRAPE1, GRAPE3) are low precision boards (18 bit or $\approx 1\%$ accuracy in the force), which are sufficient for most astrophysical applications. The machines with even numbers (GRAPE2 and GRAPE4) are working with 32 and 64 bit arithmetic and are designed to calculate molecular systems and specific stellar dynamical problems. In GRAPE3, one GRAPE1A board is put into a customized LSI chip. Presently, GRAPE3Af, which consists of 8 such chips, is produced in a small series and is available for about US $ 20000. Its peak performance is 4.8 Gflops. Up to 16 boards can be put together to work in parallel. Up to 1995 the GRAPE4 project should be finished. In GRAPE4, the GRAPE2 board is put into a LSI chip and 1500 of such chips are combined together. This board will reach a performance in the Teraflop regime. Although the flop rate of GRAPE is very impressive, one should keep in mind that PP techniques have a much larger operation count to calculate the force on a particle than the approximative techniques mentioned above. Furthermore, in hydrodynamic simulations, a non–negligible part of the computational time is necessary to calculate the pressure force and the equation of state.

In a series of publications the Tokyo group has shown that it is possible to perform large N–body simulations on such a board with a speed close to its peak performance. Furthermore, a SPH code was implemented. In such a code, the gravitational force and the neighbour list was obtained with GRAPE, the evaluation of the hydrodynamic force and the solution of the equation of state being done on a workstation. A tree code was implemented on GRAPE, too. Again, the force evaluation is done on the board, but the tree construction and the determination of the interaction list has to be done on the host. Thus, a powerful workstation is essential, in order to use a SPH and/or a tree code with GRAPE efficiently. It should also be no serious problem to implement a $P^3M$ on GRAPE: The PM part is done on the host, the PP part on GRAPE.

## 4 Results

The following comparison holds for a multiple timestep SPH–N–body tree code [3] written and optimized to run on a CRAY. All CPU timings are given for one processor of a CRAY YMP 4/64 (333 Mflops). The timings for GRAPE are obtained on one single GRAPE 3Af board (8 LSI chips, 4.8 Gflops). The host is a SPARC10 clone ($\approx$ 15 Mflops). The unchanged tree code runs about 20 times slower on the SPARC10 than on the CRAY. Using GRAPE the main code structure remains unchanged, only the subroutines for the force calculation and the neighbour list are replaced by the GRAPE routines, i.e., we compare a *tree code* on the CRAY with a *PP code* on GRAPE. To accelerate the computations on the workstation, REAL*4 arithmetic is used whenever possible. Only little effort was spent to optimize the host calculation for the SUN. Comparing N–body simulations one should keep in mind that the N–body system is chaotic. Thus, it is not possible to compare position, velocities and other properties of specific particles, but only the structure and kinematic of the whole system.

To become familiar with GRAPE and to adapt the N–body code required only two days. A 4000 (33000) body simulation requires 40 min (10 h) on the CRAY. The same result was obtained with GRAPE in 9 min (1.7 h). More than

80% of the calculations are done on the board. In both cases, the system ends in a highly clustered state, which is advantageous for GRAPE. In the case of a $64^3$ calculation, the CRAY is about two times faster for a moderately clustered system, in the case of a highly clustered system two times slower. The break even point between PP on GRAPE and tree on CRAY is of the order of a few $10^5$ particles. The tree code requires 220 MB of memory, the GRAPE only 50 MB (REAL*4), i.e., one can perform the same simulation without any problem on a mid class workstation. Running the same code on two boards gives a speed up of 1.1, 1.5 and 1.9 for 4000, 33000 and $64^3$ particles. A multi board version combined with a fast workstation would allow simulations with several million particles within a few days.

In the last paragraph we discussed the performance of GRAPE for a well suited problem. Even more interesting is its performance for more general problems, which only partially exploit the special features of GRAPE. In the CRAY code, the computing time for the SPH part is about 10%, i.e., the problem is computationally dominated by gravitation. In contrast to the previous problem, some changes in the algorithms are necessary before SPH runs efficiently, but the effort for these changes is negligible compared to the effort necessary to run the code on a parallel platform. The resulting code requires about 7 hours for a simulation with 4000 SPH and 4000 dark matter particles. This is about 1.3 times faster than on the CRAY. About 80% of the computations were done on the workstation. Replacing the relatively slow SPARC10 by a faster one, a speedup of a factor 2 or even more should be easily possible. The behaviour should be even better for larger particle numbers, because the performance on GRAPE is limited by the SPH part, which grows $\propto N$, whereas the performance on CRAY is limited by the gravitational force calculation which grows like $N \log N$. Furthermore, judging from our experience the maximum possible simulation on GRAPE will not be limited by the $N^2$ behaviour of GRAPE for large particle numbers but rather by a shortage of memory and I/O operations on the workstation. Therefore, we believe that it makes no sense to use more than one board for such simulations. In another simulation, we have taken star formation into account, i.e., in gravitationally unstable regions, some mass is removed from the gas particles and put into new collisionless star particles. Consequently, during a simulation the number of N–body particles continuously grows. Typically, 10000–25000 star particles are formed from 4000 gas particles [14]. Because all new particles are located in the densest regions, the degree of clustering increases, which results in a larger computing time on a CRAY (about 60 hours). On GRAPE the computing time is still limited by the hydrodynamic part, which does not depend on the number of star particles. Therefore, the computing time grows only moderately to about 20 hours, i.e., GRAPE is three times faster.

## 5 Conclusion

We have presented a highly active and interesting field of current astrophysical research, namely structure formation in the universe. We have shown that the scientific progress in this field is tightly coupled to the capabilities of the available supercomputers. As an alternative, a combination of a workstation with the special purpose hardware GRAPE can solve problems of comparable size within a time comparable to that on a supercomputer like a CRAY. The GRAPE hardware is applicable to N–body simulations and SPH hydrodynamical simulations the power of the host being crucial for a good performance in the latter case. A meaningful use of GRAPE requires a problem dominated by the calculation of gravitational forces. The force evaluation on the low precision boards (GRAPE3) is only accurate to $\approx 1\%$. Numerical simulations which demand a higher precision should be run on the high precision boards (GRAPE4). In that case, however, an investment of about US $ 100 000 is necessary to get a similar performance. In summary, the GRAPE hardware is applicable to a whole class of astrophysical problems. It is a very attractive alternative to current supercomputers, at least for problems which are dominated by CPU time rather than by memory.

**Acknowledgements:** I would like to thank D. Sugimoto for the possibility to use the GRAPE board in Tokyo. I. Hachisu, J. Makino and M. Taiji are gratefully acknowledged for their help during my work on GRAPE. E. Bertschinger, E. Müller, R. Spurzem and S. White are greatfully acknowledged for discussions and suggestions about the different aspects of large scale N–body and hydrodynamical simulations. All CRAY simulations have been performed on the CRAY YMP 4/64 of the Rechenzentrum Garching and all GRAPE calculations on the GRAPE3Af board of the University of Tokyo.